\documentclass[aps,prl,twocolumn,showpacs,superscriptaddress,groupedaddress]{revtex4}  
\usepackage{graphicx}  
\usepackage{dcolumn}   
\usepackage{bm}        
\usepackage{amssymb}   
\usepackage{color}

\usepackage{amsmath, amsfonts}
\usepackage{nccmath}
\usepackage{etoolbox}

\usepackage{gensymb}
\begin{document}

\title{Universality in the initial electrocoalescence}

\author{Juan S. Marin Quintero}
\author{Markus C\"asar}%
\author{Prashant R. Waghmare}
 \email{waghmare@ualberta.ca}
\affiliation{%
 interfacial Science and Surface Engineering Lab (\textit{i}SSELab), Department of Mechanical Engineering, University of Alberta, Edmonton, Canada\\
}%

\date{\today}

\begin{abstract}
The characteristic time for the electrocoalescence of two sessile drops is identified and verified for wide range of operating parameters such as Ohnesorge, Electrowetting number, driving frequency and drop to surrounding medium viscosity. Elegant mechanism is devised to circumvent the repulsion and promote the coalescence at the merging three phase contact lines of two electrospreading drops. With the revised definition of characteristic time scale, we demonstrated the universality of the bridge growth at the early times of coalescence for more than twenty different scenarios in triplicates. The universal bridge growth can only be obtained with the consideration of proposed time that considers the time to reach the maximum deformation in the drop, which is always ignored while defining the characteristics time scale of coalescence. 
\end{abstract}

\maketitle
One dimensional lubrication approximation is commonly adopted approach for theoretical analysis of the bridge growth observed during the coalescence. The scaling efforts are nonunanimous, the bridge growth was reported to be linear~\cite{Snoijer_PRL_2012,mitra2015symmetric}, squared~\cite{Aarts_2005_PRL} and even cubed~\cite{lee2012coalescence,Ristenpart2006,diez2002computing} for the very same sessile drop coalescence process. With the consideration of lubrication approximation along with the Tanner's law~\cite{Ristenpart2006,lee2012coalescence} ${1/3}^{rd}$ scaling is also witnessed as observed here in the case of electrocoalescence. Electrocoalescence, is the merging of two liquid drops under the influence of an electrical field~\cite{Stone_Nature_2009}. Here, we proposed the scaling argument for the electrocoalescence of two spreading droplets, that can be extended to sessile droplet coalescence scenario. The universality in the characteristic time and corresponding bridge growth dynamics is established for wide variety of electrospreading parameters along with the drop and surrounding medium properties.
 
Drop coalescence can be from suspended droplets~\cite{eggers1999coalescence}, hanging droplets~\cite{Basaran_PNAS_2012}, and or spreading droplets~\cite{Snoijer_PRL_2012}. Moreover, it can be in the presence or absence of electric, magnetic or acoustic field. The spreading drop and hanging drop coalescence have always been surface tension dominant event. Interfacial and the thermophysical properties dictate the bridge growth that is formed between two droplets and subsequently the mass transfer rate. The characteristic capillary velocity and time govern the bridge growth and mass transfer dynamics~\cite{eggers1999coalescence,Aarts_2005_PRL,wu2004scaling}. Mass transfer rate, triggered due to capillarity, is always constrained by the properties of systems which has been overcome by imparting external fields. Actuation of electric field to control the rate of coalescence has always been one of such approach but interestingly enough only studied for suspended~\cite{atten2006simplified} or hanging droplets~\cite{lundgaard2002electrocoalescence}. Inability to control over the approach speed of the three phase contact line (TPCL), in turn the bridge growth, might be the source for this discrepancy in the literature. The challenge of maneuvering the drop spreading velocity can be circumvented by means of electrowetting. Here, a detailed study for a wide variety of condition for electrocoalescence is presented and the one-third law for bridge growth is observed for all studied cases which was further corroborated with a scaling analysis.
\begin{figure}
\centering
\includegraphics[width= 8.2cm ]{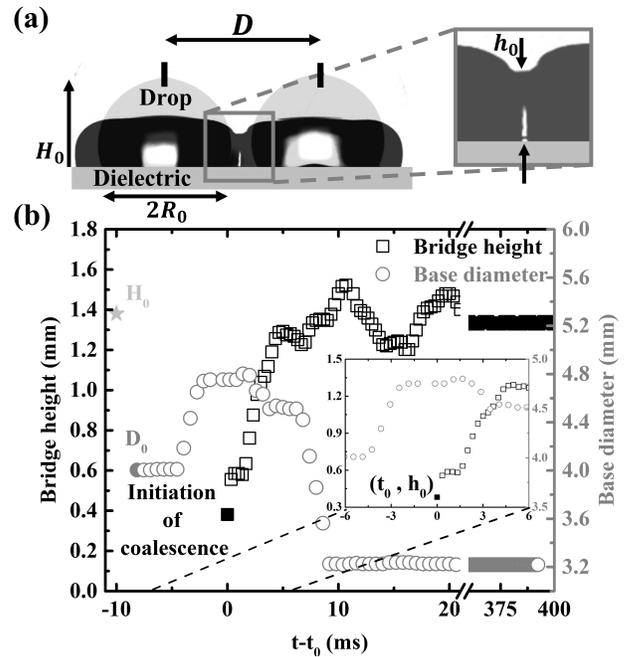}
\caption{Bridge growth response of two drops induced by a single wave of electrowetting (a) Schematic of two drops on a dielectric layer with a conductor wire at the top, where the initial bridge height is detailed as depicted in the inset. (b) Temporal variation in the bridge height and base diameter for coalescing drops with a viscosity of $0.91\, mPa\cdot s$ and separated approximately by $2.4\, mm$ centre to centre, submitted to a wave of $350\, V$ at $100\, Hz$.}
\label{Figure_1}
\end{figure}
For electrocoalescence of two spreading droplets, it was articulated that the presence of similar kind of charges at the TPCL restricts the drop coalescence at the TPCL~\cite{Stone_Nature_2009,quilliet2001electrowetting,bansal2018effect}. A unique way was proposed to minimize this repulsion and obtain the coalescence closer to TPCL. As shown in Fig.~\ref{Figure_1}(a), two equilibrated sessile droplets, separated by a finite distance $(D)$, are forced to spread by electrowetting. A device is engineered to maintain precise separation distance $D$ between two electrodes and the drop volume is adjusted to assure the absence of repulsion at the interfaces. The engineered device can be seen at the supplemental material Fig.S1~\cite{supplementary_material}, along with its details and geometric calculations to determine the appropriate separation distance based on the drop volume. Prior to the interactions of two TPCLs, after the actuation for electrospreading, both the drops detach from the needles and electrodes simultaneously. This not only confirmed the equal spreading rate but assured the mass transfer is through TPCL only. The merging of drops along the drop-medium interfaces is meticulously avoided by selecting the appropriate separation distance and drop volumes. Bridge growth $(h_0)$ dynamics was captured by a high speed camera (Nova S9) at 15,000 fps with a spatial resolution of $10\, \mu m$/pixel. A copper substrate coated with a $20\, \mu m$ PDMS layer, acting as a dielectric for the electrowetting process, is used for all the cases. The drops consisted in aqueous solutions of glycerol and $0.1\, M$ sodium chloride. To control speed of the merging TPCL and subsequent bridge growth dynamics, the applied voltage, frequency, and viscosity of the drop and medium were changed. 

Before the coalescence starts, the base diameter of spreading drop momentarily attains a constant magnitude and eventually the bridge growth starts, refer to Fig.~\ref{Figure_1}(b). At the onset of coalescence, the bridge grows rapidly with no change in the base diameter. As the bridge height surpasses the initial equilibrium drop radius, the drop experiences the pull at the noncoalescing end due to sudden flux towards the merging side of the droplets. This results in the decrement of the base diameter. Despite the oscillatory growth in the middle, the non-coalescing ends are free from such oscillations. This can be attributed to the pinning of the contact line and momentum transfer through the drop and medium. The intermittent constant base diameter phase is observed that can be due to the hysteresis experienced by the moving TPCL.

\begin{figure}
\centering
\includegraphics[width= 8.2cm ]{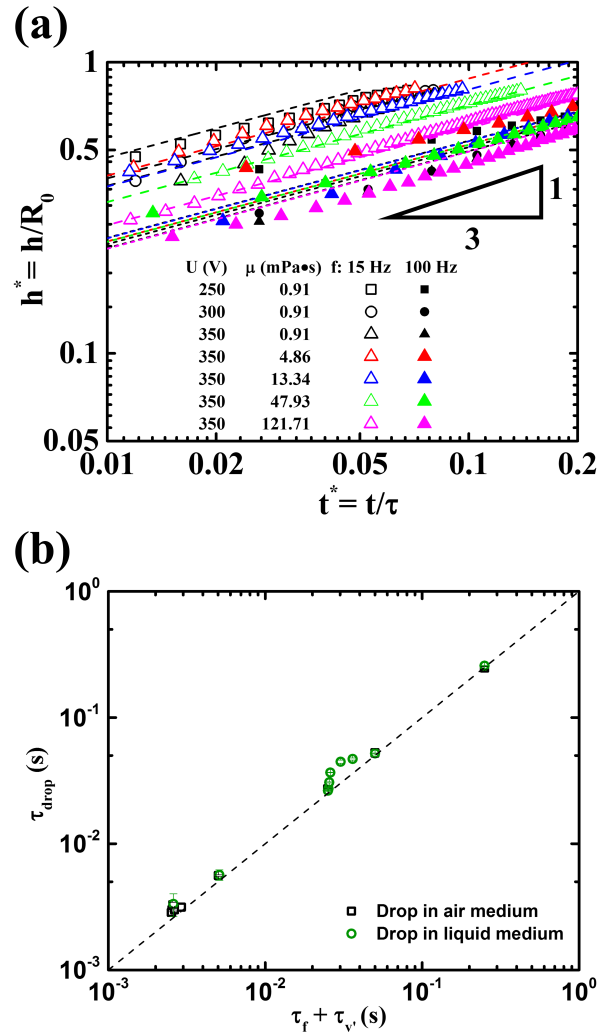}
\caption{Universal response of drops to electrowetting. (a) Universal non-dimensional bridge growth of drops following a $1/3$ power law for different voltages, viscosities, and frequencies. (b) Universal time scale of electrowetting drops for attaining the maximum spreading as a function of the frequency and viscous time scale. }
\label{Figure_2}
\end{figure}

With an additional driving and resisting force due to applied voltage and surrounding medium, respectively, the revised scaling argument for the bridge growth is propped, as presented here: 

\useshortskip
\begin{center}
\begin{equation} 
h^* \sim \left(\frac{ t}{ {\frac{1}{4f}}+(\mu\mu_0)^{1/2}(R_{max}-R_0)/\eta\gamma}\right) ^{1/3}
\label{eq1}
\end{equation}
\end{center}
\noindent
In Fig.~\ref{Figure_2}, the bridge growth in the presence of liquids with viscosity that varies from $0.91$ to $121.1\, mPa\cdot s$ is presented; the similar combination of viscosities is also investigated for varied electrowetting parameters, i.e., electrowetting number $(\eta)$ and frequency of applied voltage ($f$). As depicted in Fig.~\ref{Figure_2}(a), the nondimensionalized temporal $(t^*)$ variation in the bridge height $(h^*)$ growth is presented. Here, height is normalized with the initial drop radius $(R_o)$ and the characterize time, $\tau$, is used to obtain the nondimensional time, $t^*$. The $\tau$ is combination of the maximum spreading time which is a function of the actuation time of the applied frequency ($\tau_f$) and the modified viscous time scale ($\tau_v'$). It is to be noted that the interfacial tension, $\gamma$, for each scenario was in the range of $65.6 - 71.7\, mN/m$. Experimental results of at least fourteen cases with triplicates, denoted by symbols in the Fig.~\ref{Figure_2}(a), clearly follows Eq.~\ref{eq1}, justifying the revision proposed to the well accepted scaling argument for droplet coalescence~\cite{Ristenpart2006,lee2012coalescence}. 

The necessity to consider the $\tau_f+\tau_v'$ as a characterize time ($\tau$) is presented in Fig.~\ref{Figure_2}(b). The longest time required for the drop to attain maximum deformation, in response to the applied perturbation, is considered as $\tau_{drop}$. For a sine wave actuation, caused by AC signal, the maximum voltage occurs at the quarter of the cycle. In an idealistic scenario, in the absence of influence of any thermophysical properties of drop and medium, $\tau_{drop}$ must be equal to $1/4f$. However, the drop has a different response time based on the system's thermophysical properties and this $\tau_{drop}$ can be either inertial, viscous, or capillary time scale~\cite{das2012early}. The viscosity of the surrounding medium is varied in three orders of magnitude, therefore, viscous time scale was the first choice to be characteristic time. We can contemplate that the maximum force per unit contact length, exerted at the TPCL, is countered by a resisting offered by the viscosities of drop and the surrounding medium \cite{vo2018contact}. Thereafter, the balance of these two forces will lead to a characteristic velocity that governs the motion at the TPCL between the domain of maximum ($R_{max}$) and minimum spread ($R_0$) of the drop. The resultant viscous time scale, considering the role of surrounding medium and electro-spreading, can be deduced as $\tau_v'=(\mu\mu_0)^{1/2}(R_{max}-R_0)/\eta\gamma$, where $\mu$ and $\mu_0$ are the viscosity of the drop and the medium, respectively. Due to the diffusion delay, the drop response will always be greater than imposed actuation time, yet it can be delayed by the different factors as noticed in this study. If viscous time scale is considered as an additional time required to transfer the momentum, the total response of the drop ($\tau$) is the sum of $\tau_f$ and $\tau_v'$, and Fig.\ref{Figure_2}(b) depicts validity of this approach. The drop response time is dominated by the actuation time of the external force, with marginal influence from the applied voltage and the viscosity of both drop and liquid medium. Based on this detailed investigation, in the case of electrocoalescence analysis we have utilized $\tau$ as characteristic time for the analysis. One can argue the validity of such a revised scaling argument without any external actuation or surrounding medium, therefore, the scaling law is validated for sixteen literature data set and the details of this comparison is presented in Fig.S3 in the supplemental material~\cite{supplementary_material}.

\begin{figure}
\centering
\includegraphics[width= 8.2 cm ]{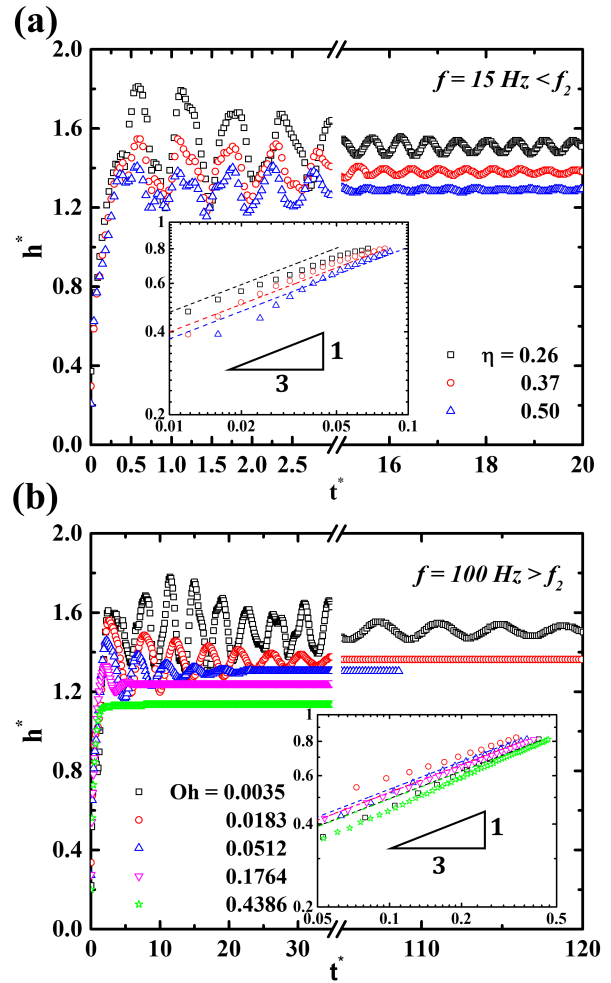}
\caption{Effect of electrowetting and Ohnesorge number on the bridge growth for drops in air separated by a distance $D/(2R_0) = 1.15$. (a) Electrowetting number effect on bridge growth for applied frequency under the resonance frequency for drops with $Oh = 0.0035$, and (b) Ohnesorge number effect on bridge growth for applied frequency above the resonance frequency for an applied wave of $\eta =0.5$. The initial bridge growth is presented in the inset figures.}
\label{Figure_3}
\end{figure}

{\it{Role of external actuation}}: Applied electric field is the actuation force that provokes the drops to spread and coalesce in desired fashion. The magnitude of this actuation force can be estimated based on the electrowetting number $\eta=\varepsilon_0 \varepsilon_d U^2/(2d\gamma)$. For a fixed dielectric constant $\varepsilon_d$, dielectric thickness $d$, and marginally changing surface tension $\gamma$ scenario, the applied voltage characteristics, i.e., it's magnitude $(U)$ and frequency $(f)$ dictate the bounding limit of $\eta$. The lowest possible limit that initiated the coalescence is $\eta=0.26$ and the contact angle saturation~\cite{verheijen1999reversible} restricts the variation until $\eta=0.50$. The constrain for $\eta$ in relationship with separation distance between the drops can also be established as presented in the Fig.S2 in supplemental material~\cite{supplementary_material}. The resonance frequency of the drop is another constrain~\cite{oh2008shape,hong2012frequency} for the actauation. The first associated resonance frequency ($f_2$) for a water drop of $3\, \mu L$ is approximately $f_2\approx 40\, Hz$. Therefore, in order to avoid the influence due to the closeness to the resonance frequency, the bridge growth was studied for two applied frequencies ($15\, Hz$ and $100\, Hz$), which are under and above the resonance frequency, respectively. With these bounding limits for operating electrocoalescence, the validity of the Eq.~\ref{eq1} is also presented for varied $\eta$ and $Oh$ in Fig.~\ref{Figure_3}.

The oscillatory behaviour in the bridge growth for both the frequencies is noticed in every case except for very high viscous drops scenario $(Oh=0.438)$ as depicted in Fig.~\ref{Figure_3}. For the brevity purpose, the variations in $\eta$ and $Oh$ is presented for below and above the resonance frequency, respectively. Observations for frequencies above the resonance frequency with varied $\eta$ is similar to the Fig.\ref{Figure_3}(a) and in the case of $Oh$, the bridge growth behaviour is also independent of frequency. One-third scaling argument presented in Eq.~\ref{eq1} remains valid for these cases as well. Thus, one can argue that the proposed modifications that accounts actuation characteristics and surrounding parameters are paramount to get the appropriate scaling law. The initial bridge growth, until it attains the first maxima, follows the universality irrespective of operating parameters, i.e., for $\eta$, $Oh$, and $f$. 

Coalescing and oscillatory motions are dependent on the operating parameters. Since $\eta$ is the driving force that dictates the maximum spread, at lower $\eta$, the higher velocity is evident that resulted in long lasting momentum in decayed oscillatory bridge growth. Interestingly, the oscillatory behaviour in the bridge growth resembles the inertial oscillatory motion~\cite{shardt2014inertial}. At certain growth rate, the induced force at the TPCL during the coalescence is sufficient to overcome the adhesion due to the contact angle hysteresis and to slip the TPCL. This slip is the indication of a change in the base diameter, as detailed in Fig.\ref{Figure_1}(b). One can completely eliminate this slip-stick motion by performing experiments on a substrate without contact angle hysteresis, but it is challenging in air medium. Nevertheless, one can argue that if the actuation frequency is kept high enough, to ascertain the higher force at TPCL to overcome this hysteresis. By modifying the viscosity of the drops, one can study the role of $Oh$. Higher $Oh$ implies a stronger effect of the viscous forces that decelerates the bridge growth \cite{mitra2015symmetric,Snoijer_PRL_2012}. Also, the increment of $Oh$ implies a lower spreading velocity and longer time required for the drop to respond to the external actuation force and to restructure. Hence, this delay will prompt smaller perturbations in the drop during the coalescence as observed in Fig.\ref{Figure_3}(b). The eventuality of the oscillatory motion and the magnitude of the maximum bridge height becomes less pronounced as $Oh$ is increased. The change in the magnitude of the bridge height and the frequency of oscillatory motion is associated to the underdamped regime, in which the drop oscillates and the magnitude of these oscillations decay exponentially over time \cite{lin2018impact}. The deformation of the drops in the overdamped regime did not show capillary waves, and the time to reach the maximum bridge height was delayed \cite{thoroddsen2005coalescence}. An additional effect is witnessed for high $Oh$, where the base diameter of both drops was completely pinned during the coalescence time. Thus, the base diameter of the newly formed drop is larger with lower equilibrium height. 

In the context of drop coalescence with controlled external force like electrowetting, magnetowetting, and acoustic-wetting, it is paramount to include the necessary operating parameters in the scaling analysis which can obey the experimentally observed coalescing behaviour. We demonstrated for air and liquid medium droplet spreading due to AC electrowetting, the frequency of the applied actuation is key parameter in characteristic time. And finally, with the appropriate definition of the characteristic time scale, the universal response in the bridge growth of two electro-colescing droplets is observed, which follows a $1/3^{rd}$ power law. 

We thank Natural Sciences and Engineering Research Council (NSERC) for the funds with grant number RGPIN-2015-06542. We also thank Hafijur Rahman for our fruitful conversations regarding the scaling analysis.


\end{document}